\documentclass[journal]{IEEEtran}

\usepackage[T1]{fontenc}
\usepackage{multirow}
\usepackage[table,xcdraw]{xcolor}
\usepackage{graphicx}
\usepackage{amsmath}
\usepackage{amsfonts}
\usepackage{algorithm}
\usepackage{amsthm,amssymb}
\usepackage{mathrsfs}
\usepackage{algorithmic}
\usepackage{makecell}
\usepackage{verbatim}
\usepackage{booktabs}
\setcellgapes{3pt}
\usepackage{amsthm,amsmath,amssymb,lipsum}
\usepackage{mathrsfs}
\setcellgapes{3pt}
\usepackage{multirow}

\usepackage[justification=centering]{caption}
\usepackage{color}
\usepackage[subfigure]{tocloft}
\usepackage{subfigure}
\usepackage{tabularx}
\usepackage{amsmath, bm}
\usepackage{authblk}
\usepackage{amsmath}
\usepackage{multirow}
\usepackage{tikz}
\usetikzlibrary{shapes.geometric, arrows, positioning, fit, calc}
\usepackage{bm}
\makeatletter

\newcommand{\Rmnum}[1]{\expandafter\@slowromancap\romannumeral #1@}
\makeatother

\usepackage{mdframed}
\usepackage{listings}
\usepackage{xcolor}

\mdfdefinestyle{promptbox}{
    linecolor=black,
    linewidth=1pt,
    frametitlerule=true,
    frametitlebackgroundcolor=gray!20,
    innertopmargin=10pt,
    innerbottommargin=10pt,
    innerrightmargin=10pt,
    innerleftmargin=10pt,
    skipabove=10pt,
    skipbelow=10pt
}

\usepackage{hyperref} 
\hypersetup{
	hidelinks
}
\ifCLASSINFOpdf
\else
\fi
\usepackage{hyperref}

\begin{document}

\title{From Traditional Automation to Embodied Wireless Intelligence: Vision-Language-Action Empowered Physics-Aware Communication Networks}

\author{
\thanks{Corresponding author: Kezhi Wang}
Genze Jiang, Kezhi Wang, \textit{Senior Member, IEEE}, Xiaomin Chen, Yizhou Huang
	\thanks{
    Genze Jiang, Kezhi Wang and Yizhou Huang are with the Department of Computer Science, Brunel University London, UK (e-mail: Genze.Jiang@brunel.ac.uk, Kezhi.Wang@brunel.ac.uk, Yizhou.Huang2@brunel.ac.uk).
    
    Xiaomin Chen is with the Department of Computer Science, University of Reading, UK (e-mail: xiaomin.chen@reading.ac.uk)
    }
}

\markboth{Submitted for Review}%
{Shell \MakeLowercase{\textit{et al.}}: Bare Demo of IEEEtran.cls for IEEE Journals}
%

\maketitle

\begin{abstract}
Wireless network automation has progressed from rule-based self-organising networks (SON) to data-driven optimisation, yet existing systems remain fundamentally disembodied. They act on performance indicators without perceiving the physical environment that governs radio propagation. We propose the embodied intelligent empowered base station (eBS), a paradigm that adopts a Vision-Language-Action (VLA) pipeline to transform base stations into autonomous agents capable of situated perception, causal physical reasoning, and physics-aware action generation. The eBS employs a two-tier asynchronous architecture: a Semantic Planner powered by a frontier Vision-Language Model (VLM) generates structured action directives on human timescales, whilst a Tactical Controller executes real-time adaptation. Case studies demonstrate that a single VLA pipeline, without task-specific training, can perform zero-shot material reasoning, generalise across viewpoints, and predict dynamic events before signal degradation occurs. These results illustrate a paradigm shift from traditional rule-following network automation to embodied intelligence empowered future wireless networks.

\end{abstract}


\section{Introduction}

Network automation has been a design goal of cellular systems since the 4G era. Self-organising networks (SON) configure, optimise and heal the network through predefined rules and threshold triggers, and machine learning has since replaced many of those rules with learned models \cite{klaine2017survey}. Both families share one limitation. They act on performance indicators such as throughput, reference signal received power (RSRP) and cell load, but they never observe the physical environment that produces those indicators. This article argues that the missing ingredient is embodiment, and that the Vision-Language-Action (VLA) paradigm now established in robotics can provide it.

\subsection{From Traditional Automation to Embodied Intelligence}

Early industrial robots executed fixed, preprogrammed procedures. An arm repeated one trajectory regardless of workpiece variation. A pick-and-place system moved to predetermined coordinates. These machines achieved automation but not intelligence. They could not perceive their environment, reason about unexpected situations, or adapt to novel conditions.

Embodied intelligence has since transformed robotics. Modern robots run on foundation models such as VLA architectures \cite{zitkovich2023rt}. They observe their surroundings through cameras, reason about objects and spatial relationships in natural language, and generate actions grounded in physical understanding \cite{liang2023code}. Systems that follow rules have been replaced by agents that understand the world.

Wireless network management has followed a parallel trajectory, but it has stalled before the final step. SON algorithms execute predefined policies whenever a key performance indicator crosses a threshold \cite{klaine2017survey}. Machine learning optimisers learn statistical correlations between network metrics, and they learn nothing about the physical world that generates those metrics. Radio access network intelligent controllers (O-RAN RIC) and AI-native architectures have extended this automation without changing its character. A base station equipped with such systems detects that signal quality has degraded. It cannot \textit{perceive} the physical cause, \textit{reason} about the electromagnetic properties of the surrounding materials, or \textit{predict} that a moving object is about to block a propagation path. Today's ``autonomous'' networks therefore automate reactions to symptoms, and they diagnose no causes.

The insight motivating this work is simple. The paradigm that transformed robotics can close the same gap in wireless communication. A base station (BS) that perceives, reasons about, and acts on its physical environment through a foundation model can move beyond rule-following automation, and it can then observe, understand and continuously optimise its own behaviour.

\subsection{Vision-Language-Action (VLA) Paradigm}

Vision and language models have already reached wireless communications, but the existing work falls short of embodiment. Supervised deep learning trains task-specific models on paired sensor-channel data, for example a network trained to predict blockage and trigger handover from base station camera imagery \cite{charan2021vision}. Such models are pattern matchers. They cannot generalise beyond their training distribution, explain their decisions, or transfer to a new deployment without retraining. More recent work embeds VLMs and large language models (LLMs) into wireless systems through contrastive learning frameworks \cite{wang2025vision}, fine-tuned multimodal beam predictors \cite{zhao2025multi}, vision-empowered LLM beam prediction \cite{zheng2025beamllm}, LLM-based network agents confined to the text domain \cite{tong2025wirelessagent}, and large multimodal models used as feature extractors \cite{xu2024large}. Every one of these systems still fine-tunes or trains a model for the wireless task, and none of them closes a perception-reasoning-action loop. The foundation model sits inside a conventional pipeline as a processing component, and it never acts as an agent. The eBS differs on both counts. It invokes an off-the-shelf frontier VLM with no wireless-specific training, and it lets that model issue the control action.

Robotics has already built the missing piece. Prompting a general-purpose language model to emit structured action representations, such as executable code or JSON commands, lets an embodied agent perform complex tasks with no task-specific training \cite{liang2023code}. VLA models later formalised this approach and unified visual perception, language-based reasoning and action generation in one architecture \cite{zitkovich2023rt}. Wireless researchers have noticed the opportunity. AI embodiment has been proposed as an organising principle for 6G \cite{bariah2024ai}, and large models have been proposed as the cognitive core of an integrated perception, communication and computation network \cite{li2025large}. Both proposals remain at the level of architectural vision. Neither shows how environmental reasoning becomes a physical-layer control action. This article supplies that missing step and proposes a complete VLA pipeline for autonomous BS operation.

\subsection{Contribution: Embodied Intelligence Empowered BS (eBS)}

We propose the embodied intelligence empowered base station (eBS), a paradigm in which a frontier VLM serves as the perception-reasoning-action engine of the physical layer. Three properties separate the eBS from both SON and existing data-driven controllers.

\begin{itemize}
    \item \textbf{Situated perception.} The eBS builds a semantic description of its local environment from multimodal sensing. It names objects, materials, spatial relationships and trajectories in natural language, and it does not reduce sensor data to isolated feature vectors.
    \item \textbf{Causal reasoning.} The eBS infers physical consequences from what it sees, without running an electromagnetic simulation. It reasons about how a material attenuates a signal, how geometry constrains coverage, and how a moving object will alter the channel. The reasoning is a chain of thought grounded in pre-trained world knowledge.
    \item \textbf{Autonomous action.} The eBS turns that understanding into a structured control action. It selects a transmission strategy from an identified material, and it initiates handover from a predicted trajectory. SON needs a rule for every anticipated scenario. A supervised model needs labelled data for every new deployment. The eBS reasons from first principles, acts in situations it has never seen, and accepts operator guidance in natural language.
\end{itemize}

Three limitations of the existing literature motivate this design, and we return to each of them in the body of the article. Existing VLM and LLM integrations use the foundation model as a feature extractor rather than as an agent. The closest embodied proposals stop at architectural vision. Current vision-aided systems are black boxes, they offer the operator no interpretable reasoning, and they still require task-specific training for each deployment.

One engineering obstacle stands in the way. A VLM needs 1--2 seconds to reason about an image, whilst physical-layer control should respond inside a 10-millisecond radio frame. The eBS resolves the conflict with a two-tier asynchronous architecture that decouples strategic reasoning from tactical execution.

The remainder of this article is organised as follows. Section \ref{sec:ebs} presents the eBS architecture. Section \ref{sec:case_studies} reports three case studies covering material, spatial and temporal reasoning. Section \ref{sec:discussion} discusses latency, edge hardware, real-world robustness, energy and privacy. Section \ref{sec:conclusion} concludes the article.

\section{Embodied Intelligent Base Station}
\label{sec:ebs}

\subsection{Two-Tier System Architecture}

\begin{figure*}[!t]
\centering
\includegraphics[width=0.97\textwidth]{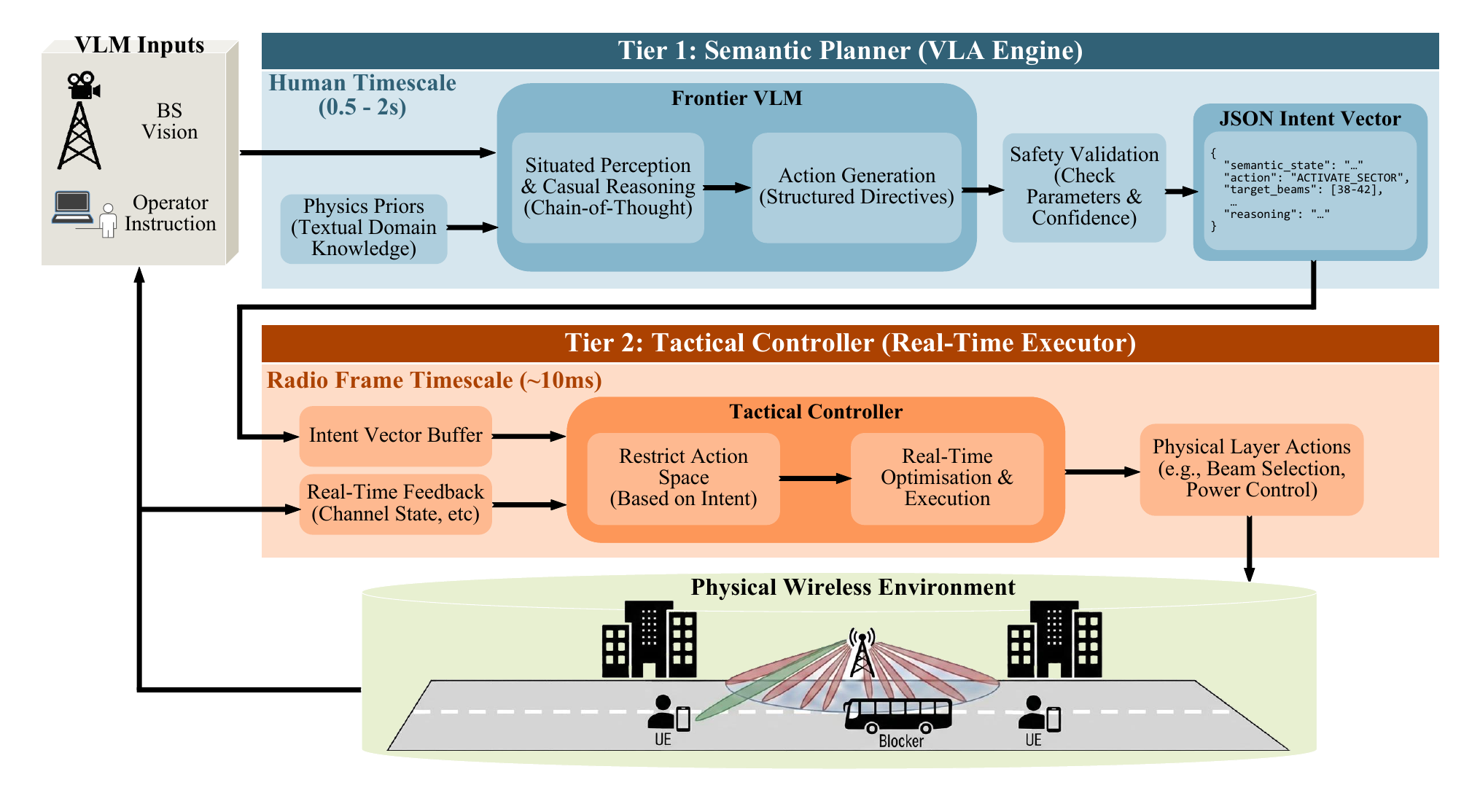}
\caption{The eBS system architecture. The Semantic Planner (Tier 1) operates on human timescales (0.5--2.0 seconds) to generate semantically informed control directives, enabling the Tactical Controller (Tier 2) to execute real-time physical-layer adaptation within the 10-millisecond radio frame constraint.}
\label{fig:architecture}
\end{figure*}

A VLA pipeline for wireless control meets a timing contradiction that robotics never faces. A robot tolerates an action cycle of 100--500 milliseconds. Physical reasoning needs a large foundation model such as GPT-4o, and inference on such a model exceeds 1 second. The radio interface allows no such budget. A vehicular mmWave link at 28 GHz has a channel coherence time below 1 millisecond, and even the beam coherence time, which measures how long a chosen beam stays aligned and is an order of magnitude longer than the channel coherence time, spans only tens of milliseconds \cite{va2017impact}. Direct integration is therefore infeasible. The environment moves out from under the VLM before the VLM finishes reading a single frame. We resolve the conflict by splitting the intelligence across two tiers, each matched to its own timescale, as shown in Fig. \ref{fig:architecture}.

\textbf{Tier 1: The Semantic Planner.} This tier embeds a frontier VLM and performs the perception and reasoning stages of the VLA pipeline. It ingests camera imagery, historical traffic patterns and operator instructions. It runs asynchronously at 0.5--2.0 Hz and computes no physical-layer parameter directly. Instead, it translates a visual observation into a structured control directive, which constrains the action space of the tier below. In beam management, an observation such as ``user in right lane'' becomes a spatial constraint on the beam search space.

\textbf{Tier 2: The Tactical Controller.} This tier is a lightweight real-time executor that runs inside the 10-millisecond radio frame. It receives an intent vector from Tier 1 and optimises only within the subset of the action space that the directive allows. In beam management, it restricts pilot transmission to a subset of beam indices. Other physical-layer functions such as power control, handover and scheduling reuse the same architecture with a task-appropriate executor. The controller can therefore track a fast-fading channel whilst it respects the strategic constraint set by the VLA pipeline.

\subsection{Information Flow and Asynchronous Coupling}

The architectural innovation lies in the asynchronous coupling between the two tiers. The Tactical Controller holds local state and keeps optimising the physical-layer configuration from real-time feedback, even while the Semantic Planner is still processing a new frame. Physical-layer control therefore never blocks on VLM inference. Three properties follow from this coupling.

\begin{itemize}
    \item \textbf{Graceful degradation.} The Tactical Controller reverts to conventional operation, such as an exhaustive sweep over all 64 beams, whenever VLM inference is delayed or the visual link is obstructed. System reliability therefore never falls below the conventional baseline.
    \item \textbf{Progressive refinement.} The Planner accumulates observations over several seconds and refines the intent vector, which narrows the action space of Tier 2 further at each update. In beam management this narrows the search sector and reduces pilot overhead.
    \item \textbf{Operator-in-the-loop control.} A network engineer can inject a natural language instruction such as ``prioritise the emergency vehicle''. The VLM converts the instruction into an immediate intent vector update, which overrides the default policy.
\end{itemize}

\subsection{Intent Vector Representation}

The action stage of the VLA pipeline needs a structured interface between semantic reasoning and physical-layer execution. We call that interface the intent vector. An intent vector is a JSON payload that converts natural-language reasoning into a mathematical constraint that a physical-layer algorithm can consume. The representation is general and can encode diverse control directives. In beam management it encodes a beam-priority mask, a reflection annotation and a validity window. The VLM emits the following directive once it identifies a user vehicle in the right traffic lane.
\begin{lstlisting}[basicstyle=\ttfamily\scriptsize]
{
  "semantic_state": "User in Right Lane",
  "action": "ACTIVATE_SECTOR",
  "target_beams": [40, 41, 42, 43, 44],
  "exclusion_zone": [0, 1, ..., 20],
  "reasoning": "Vehicle constrained to lane geometry;
                Building at 50deg provides NLoS backup."
}
\end{lstlisting}
The Tactical Controller parses this vector and transmits pilots only on indices 40--44 in the following frames, effectively reducing the search space by 92\% compared to full 64-beam sweeps.

\subsection{VLA-based Agent Design}
\label{sec:methodology}

A VLA agent for wireless infrastructure needs three elements that the robotics domain does not supply, namely an observation space grounded in RF-relevant scene semantics, an action space mapped to physical-layer control parameters, and a policy that connects visual perception to control commands. We instantiate these design choices for beam management, though the framework generalises to other physical-layer functions.

\subsubsection{Observation Space}
The agent's observation at each planning cycle comprises an RGB image from the base-station-mounted camera concatenated with a textual physics prior that injects domain knowledge absent from the VLM's general pre-training:
\begin{lstlisting}[basicstyle=\ttfamily\scriptsize]
SYSTEM: You are controlling a 6G base station at 28 GHz.
Your goal is to maintain high-SNR connectivity.

PHYSICS PRIORS:
- 28 GHz signals are blocked by Concrete/Metal (>20dB loss).
- 28 GHz signals penetrate Glass (3-6dB loss).
- Reflection off flat surfaces is viable for NLoS paths.

OUTPUT FORMAT: JSON only. No conversational preamble.
\end{lstlisting}
This hybrid observation (visual + textual) enables the VLM to apply three categories of pre-trained world knowledge to the wireless domain. The first is electromagnetic material behaviour, since conductive materials reflect whilst dielectrics permit partial transmission. The second is projective geometry, which infers 3D spatial relationships from 2D perspective cues. The third is causal reasoning, which predicts that a moving vehicle will create a future shadow region along its trajectory.

\subsubsection{Action Space}
Unlike robotic VLA agents, whose action spaces are continuous joint torques, the wireless VLA agent operates over discrete, structured control parameters. For beam management, the action space is the set of all valid intent vectors, each specifying a beam priority mask over the 64-element codebook, a transmission strategy (i.e., direct, reflect, or penetrate), and associated power and beamwidth parameters. This structured, discrete action space is inherently more constrained than that of robotic manipulation, simplifying the generation task and improving reliability. The same intent vector framework can accommodate other physical layer actions (handover directives, power allocation, scheduling priorities) by extending the JSON schema.

\subsubsection{Policy Specification via Structured Prompting}

In end-to-end VLA models such as RT-2 \cite{zitkovich2023rt}, the policy is learned implicitly through supervised fine-tuning on vision-action pairs. In our prompting-based approach, the policy is instead specified explicitly through the prompt structure, which determines how visual observations are mapped to action outputs. The policy comprises two mechanisms.

First, we enforce a chain-of-thought action derivation by requiring the VLM to populate a \texttt{reasoning} field prior to the \texttt{action} field in its JSON output. This exploits the autoregressive generation process, in which the model may first articulate its physical interpretation of the scene (e.g., ``the obstacle is glass based on visual transparency'') before committing to an action (e.g., ``PENETRATE with medium power''), thereby improving the logical consistency between perception and action.

Second, we inject task-specific observation queries that focus the agent's attention on the relevant physical dimension, as summarised in the following case studies, which will be validated in the next section.
\begin{itemize}
    \item \textbf{Material reasoning (Case Study A)}: ``Identify the blockage material and determine whether direct transmission or reflection paths are optimal at 28 GHz.''
    \item \textbf{Spatial reasoning (Case Study B)}: ``Estimate the user equipment's semantic spatial position (left, centre, right lane) and map to corresponding beam sector indices.''
    \item \textbf{Temporal reasoning (Case Study C)}: ``Analyse dynamic object trajectories and determine whether any will intersect the line-of-sight path within 500 milliseconds.''
\end{itemize}

\subsection{Closed-Loop Safety and Fallback}

A critical difference between wireless and robotic VLA deployment is the consequence of action failure. An incorrect control action causes immediate performance degradation such as throughput loss and dropped connections, whereas a suboptimal robotic grasp can be retried. The Semantic Planner, therefore, validates every generated action before forwarding it to the Tactical Controller. The parser verifies that recommended parameters fall within valid ranges (e.g., beam indices 0--63 for the beam management case) and evaluates the VLM's stated confidence level. If confidence falls below a predefined threshold (0.7 in our implementation) or the output contains malformed JSON, the system reverts to conventional operation, ensuring that performance never degrades below baseline. This closed-loop validation distinguishes the wireless VLA agent from open-loop robotic deployments, where action failures are tolerable.

\begin{figure*}[!t]
\centering
\includegraphics[width=0.9\textwidth]{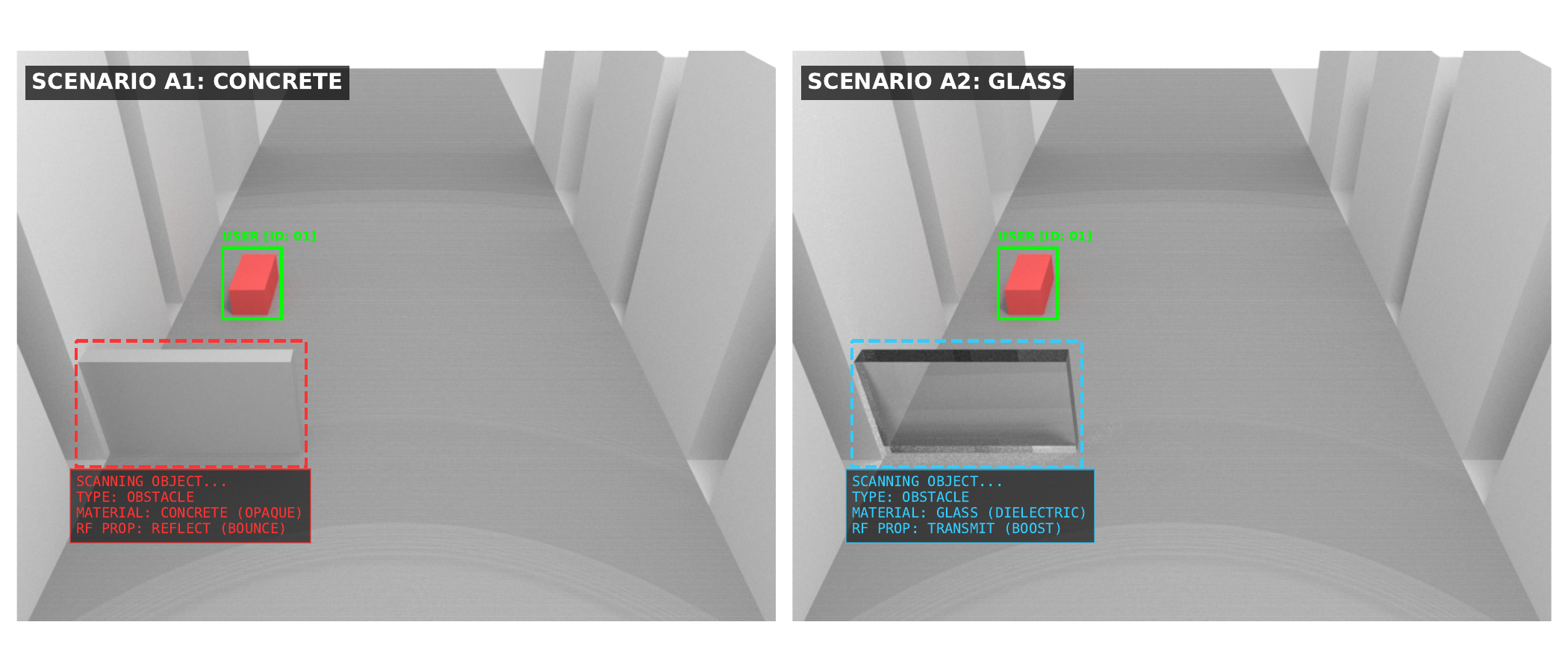}
\caption{The embodied agent identifies material properties from visual cues. Left: Detecting concrete triggers a \texttt{REFLECT} strategy. Right: Detecting glass triggers a \texttt{PENETRATE} strategy.}
\label{fig:sionna_materials}
\end{figure*}

\subsection{Multi-Station Coordination}
In conventional cooperative schemes, BS exchange raw channel state information, imposing stringent backhaul bandwidth and latency requirements that scale poorly with network density. By contrast, the eBS Semantic Planner already produces structured intent vectors that encode environmental understanding in compact, human-readable form. These intent vectors can be shared between neighbouring eBS nodes over standard backhaul interfaces, enabling semantic-level coordination. For instance, a source eBS that observes a vehicle approaching a cell boundary can transmit a trajectory prediction and material description to the target eBS, which pre-configures its beam before handover occurs. More broadly, multiple eBS nodes with overlapping fields of view can construct a shared semantic map of their environment, jointly reason about interference geometry, and coordinate coverage strategies through natural language negotiation, mirroring multi-agent VLA coordination paradigms established in collaborative robotics. This semantic coordination operates at the Planner's low update rate (0.5--2 Hz), placing negligible additional load on existing backhaul infrastructure whilst enabling a qualitatively richer form of inter-cell cooperation than numerical parameter exchange alone.

\section{Case Studies}
\label{sec:case_studies}

To validate the eBS as a VLA agent, we conduct a multi-modal experimental campaign spanning synthetic environments and controlled datasets. We utilise NVIDIA Sionna for physics-compliant ray-tracing simulations and extend validation to the Vision-Wireless (ViWi) dataset \cite{alrabeiah2020viwi} for dynamic vehicular scenarios.

A central claim of the eBS paradigm is that a single general-purpose foundation VLM can serve as a unified cognitive layer for wireless control, replacing multiple task-specific supervised models that would each require separate training data, separate architectures, and separate deployment cycles. Each case study therefore stresses a distinct capability of the VLA pipeline. \textit{Material reasoning} concerns inferring action-relevant material properties, such as reflectivity or transparency, from visual semantics to determine beam strategy. \textit{Viewpoint-invariant spatial reasoning} involves producing consistent beam predictions across novel deployment geometries without retraining. \textit{Temporal reasoning} addresses predicting future channel states from visual trajectory analysis to enable predictive handover. Together, these demonstrate that embodied intelligence emerges from the foundation model's pre-trained world knowledge rather than from task-specific pattern matching.

\subsection{Case Study A: Material Reasoning for Beam Strategy}
\label{subsec:material_reasoning}

\subsubsection{Challenge: The Semantic Gap in Channel Estimation}
Link viability at mmWave and THz frequencies depends on the electromagnetic properties of the obstacle, principally its conductivity and loss tangent, together with its thickness. Two obstacles can share a similar permittivity yet behave in opposite ways, since a thick, lossy concrete wall absorbs a transmitted path whereas a thin, low-loss glass panel passes it. Conventional channel estimators treat blockages as binary, assuming that either line-of-sight exists or it does not. This binary abstraction cannot distinguish between a concrete wall, which forces the signal onto an alternative reflective path, and a glass window, which passes the signal with only modest attenuation. This semantic gap leads to suboptimal beam selection, as the base station wastes resources attempting to penetrate opaque obstacles or fails to exploit dielectric transparency where available.

\subsubsection{Experimental Setup}
We constructed a high-fidelity digital twin using the Sionna ray-tracer. The scene depicts an urban canyon with a base station positioned at 25m height and a user vehicle at 40m distance. We introduced a variable blockage directly in the line-of-sight path:
\begin{itemize}
    \item \textbf{Scenario A1 (Concrete):} A solid concrete barrier ($\epsilon_r \approx 5.31$), thick and highly lossy, which attenuates a transmitted path beyond recovery.
    \item \textbf{Scenario A2 (Glass):} A thin glass panel ($\epsilon_r \approx 6.27$, low loss tangent), which passes a transmitted path with only modest attenuation.
\end{itemize}
The embodied agent (Tier 1: Semantic Planner) receives RGB images of the scene with the prompt: ``Analyse this base station's camera image. A user equipment is located behind the visible obstacle. Identify the blockage material and recommend a transmission strategy considering 28 GHz millimetre wave propagation.''

\begin{figure*}[!t]
\centering
\includegraphics[width=\textwidth]{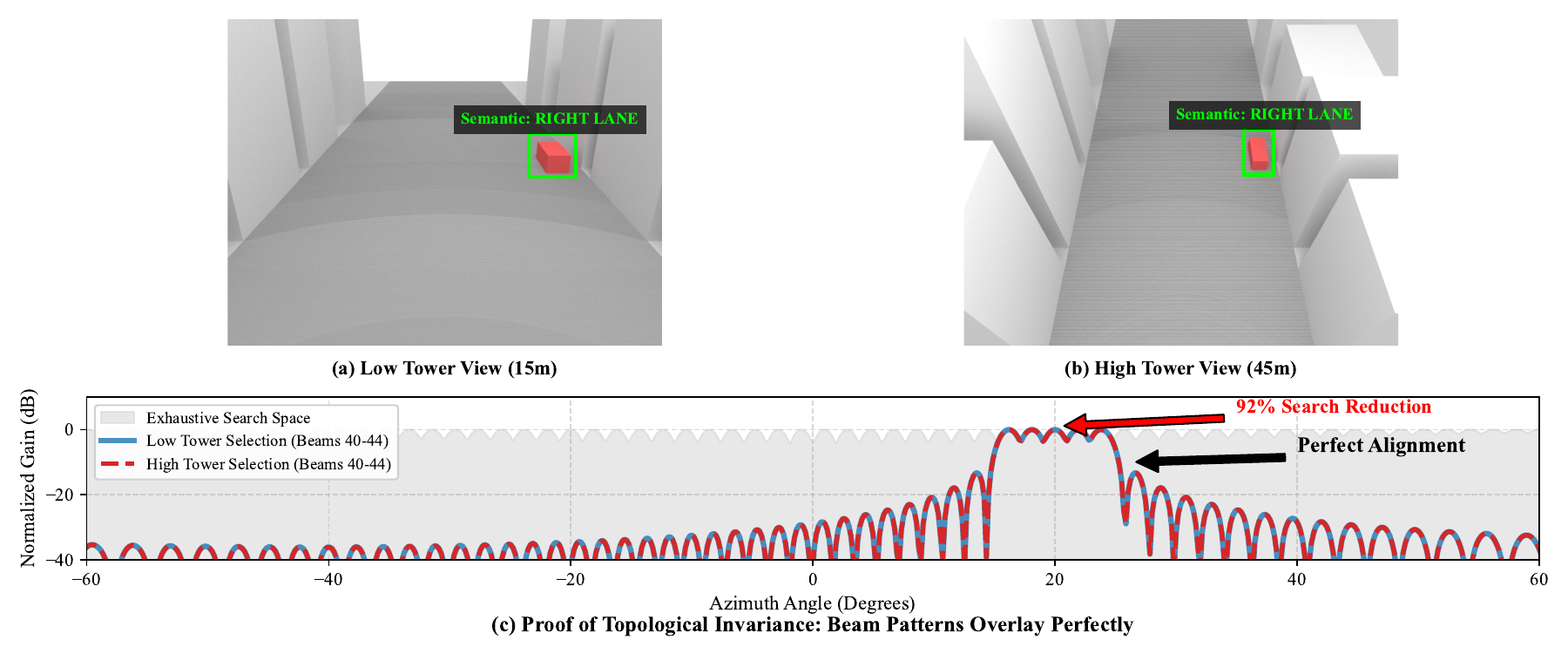} 
\caption{
(a)-(b) The agent consistently identifies the ``Right Lane'' semantic region despite a 30m height shift. 
(c) This semantic stability yields identical beam selections (Beams 40--44) for both views. The perfect overlap of the Low Tower (Blue) and High Tower (Red Dashed) patterns confirms viewpoint invariance and achieves a 92\% reduction in search overhead.}
\label{fig:generalization}
\end{figure*}

\subsubsection{Results}
The VLM successfully bridges the semantic gap through zero-shot material inference, as shown in Fig. \ref{fig:sionna_materials}.

In \textbf{Scenario A1 (Concrete)}, the agent identifies the barrier as ``opaque and solid.'' The reasoning trace states: \textit{``Reflection is necessary... a wide beam is used due to potential distortion from the surface.''} Consequently, the agent issued a directive to \texttt{REFLECT} with \texttt{HIGH} power and a \texttt{WIDE} beam width to maximise signal capture from non-line-of-sight bounces.

In \textbf{Scenario A2 (Glass)}, the agent recognises the transparency. The reasoning trace adapts: \textit{``Glass allows for penetration with slight attenuation... the view is clear, allowing for a narrow beam width for precision.''} The agent commands a \texttt{PENETRATE} strategy with \texttt{MEDIUM} power and a \texttt{NARROW} beam width to minimise interference.

\begin{table}[h]
\centering
\caption{VLM Decision Matrix for Material Reasoning}
\label{tab:materials}
\begin{tabular}{@{}lllll@{}}
\toprule
\textbf{Scenario} & \textbf{Material} & \textbf{Strategy} & \textbf{Power} & \textbf{Beam Width} \\ \midrule
A1 & Concrete (Opaque) & REFLECT & HIGH & WIDE \\
A2 & Glass (Dielectric) & PENETRATE & MEDIUM & NARROW \\ \bottomrule
\end{tabular}
\end{table}

Table \ref{tab:materials} summarises the two decisions, and both match the ground-truth optimal configuration. Sionna simulations confirm that the \texttt{REFLECT} strategy avoids a 20dB loss through the concrete, while the \texttt{PENETRATE} strategy successfully maintains a link through the glass with only 4dB attenuation. This demonstrates that the eBS can infer electromagnetic properties from visual semantics through zero-shot VLA reasoning, eliminating the latency overhead of blind beam sweeping by pre-configuring the physical layer according to actual material properties.

\subsection{Case Study B: Viewpoint-Invariant Spatial Reasoning for Beam Prediction}
\label{subsec:generalization}

\subsubsection{Challenge: Perspective-Dependent Overfitting}
Traditional supervised learning models (e.g., ResNet-based beam predictors) suffer from severe overfitting to the specific camera perspective present in training data. A model trained on imagery from a 15m tower learns that a vehicle at pixel coordinates $(u,v)$ corresponds to beam index $k$. If the camera is subsequently mounted at 45m height, the vehicle's pixel coordinates shift substantially due to the changed vanishing point, causing the model to predict incorrect beam indices. True embodiment requires \textit{viewpoint invariance}, which is the capacity to understand that a user's semantic location (and thus their azimuth angle) remains constant even when the observer's height changes.

\subsubsection{Experimental Setup}
The eBS addresses this through semantic abstraction. The Semantic Planner does not directly predict raw angles. Rather, it identifies semantic spatial regions, e.g., ``right lane'', ``intersection centre'' and maps them to physical beam sectors.

We evaluate this capability by generating two distinct viewpoints of an identical scene in Sionna:
\begin{enumerate}
    \item \textbf{Low Tower View (15m):} The camera operates close to ground level and road markings converge sharply towards the horizon.
    \item \textbf{High Tower View (45m):} The camera observes from an elevated vantage point and the scene resembles a 2D map.
\end{enumerate}

To isolate azimuthal invariance, we employ a 64-element Uniform Linear Array (ULA) generating vertically broad ``fan beams'' (elevation beamwidth $>30^\circ$) spanning $-60^\circ$ to $+60^\circ$, ensuring that users at different vertical angles remain within the same azimuthal beam index.

\subsubsection{Results}
In the \textbf{Low Tower} view, the agent identifies the vehicle in the rightmost traffic lane. Despite perspective distortion, the agent reasons: \textit{``The vehicle occupies the right lane based on road marking geometry... Azimuth angle is approximately $+15^\circ$.''}

In the \textbf{High Tower} view, despite the vehicle appearing more centrally due to the top-down perspective, the agent maintains the semantic classification: \textit{``Vehicle remains in the right lane. High-angle observation confirms lateral position. Azimuth sector remains $+15^\circ$.''}

As shown in Fig. \ref{fig:generalization}, the selected beam subsets perfectly overlapped. This confirms that the embodied agent's VLA pipeline effectively disentangles \textit{observer state} (camera height) from \textit{environment state} (user location), a hallmark of embodied spatial reasoning.

\begin{figure*}[!t]
\centering
\includegraphics[width=\textwidth]{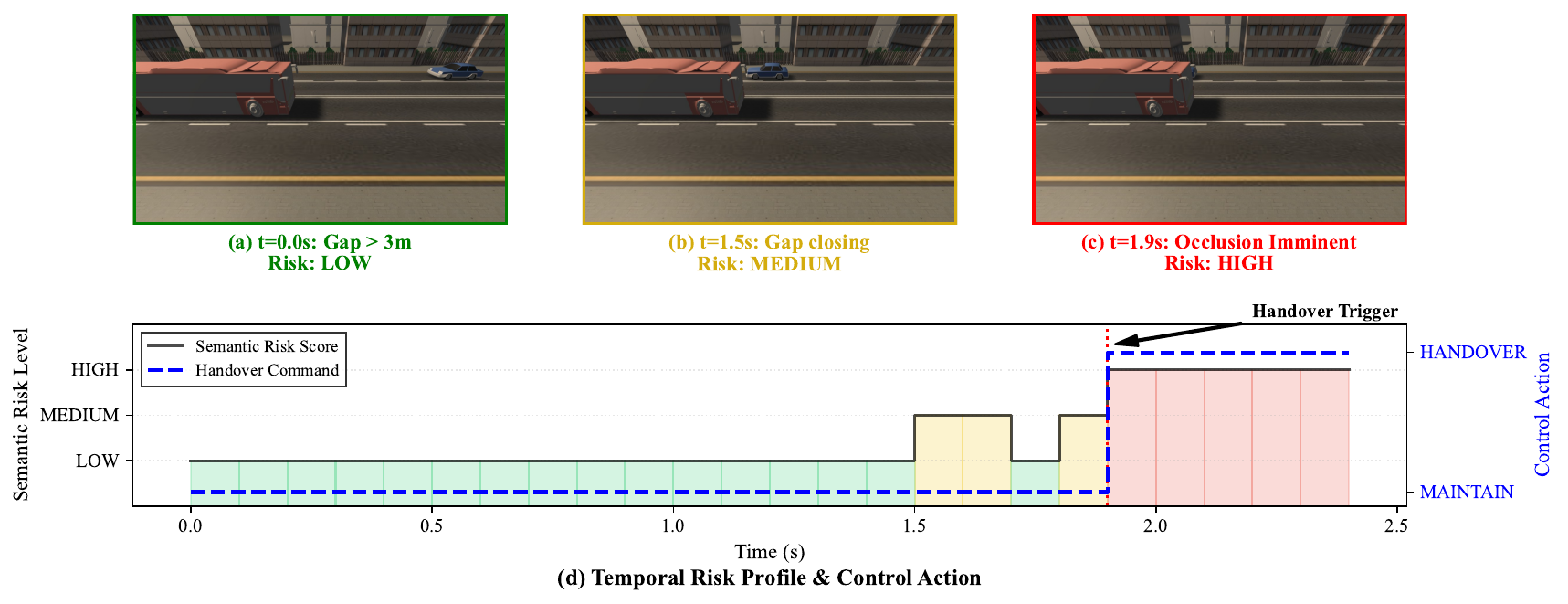} 
\caption{(a)-(c) The Semantic Planner monitors the trajectory of dynamic vehicles, elevating the semantic blockage risk score as the gap closes. (d) At $t=1.9$s, the agent predicts imminent occlusion and generates a proactive handover directive 100ms before physical link degradation occurs at $t=2.0$s.}
\label{fig:case_study_c}
\end{figure*}

Semantic guidance constrains the Tactical Controller's search space from 64 beams to approximately 5 beams (beams 40--44 for the right lane scenario). This represents a 92\% reduction in beam training overhead. Assuming 1 millisecond pilot transmission per beam, exhaustive search requires 64 milliseconds, whilst semantically guided search completes in 5 milliseconds. This efficiency gain proves critical for maintaining channel coherence in mobile scenarios.

\subsection{Case Study C: Temporal Reasoning for Predictive Handover}
\label{subsec:viwi}

\subsubsection{Challenge: Reactive Beam Management Limitations}
Standard 5G beam management operates reactively with beam search or handover procedures triggering only after Reference Signal Received Power (RSRP) degradation exceeds predefined thresholds (typically 3 dB). In dense urban environments with dynamic blockers (e.g., buses, delivery vehicles, construction equipment), this reaction latency often exceeds channel coherence time, resulting in connection interruptions and latency spikes. Recent work has demonstrated that vision-aided approaches can substantially reduce such service interruptions through predictive handover \cite{charan2021vision}. However, these methods rely on task-specific supervised models trained on paired camera-channel data. A key question for the eBS paradigm is whether a general-purpose VLA pipeline, without task-specific training, possesses \textit{temporal reasoning} capability, that is, the capacity to reason about dynamic scenes, predict future channel states from visual trajectory analysis, and generate pre-emptive control actions.

\subsubsection{Experimental Setup}
We utilise the ViWi dataset, specifically the co-located camera scenario featuring continuous image sequences of vehicles moving through dynamic street environments. The scenario comprises a serving base station on the left side of a two-lane road, with a neighbouring base station covering the opposite direction. A large bus travels in the near lane whilst the user vehicle occupies the far lane. As the user vehicle approaches the bus from behind, the bus progressively occludes all feasible beam paths from the serving base station, necessitating handover to the neighbouring base station which maintains unobstructed line-of-sight propagation.

Since the ViWi dataset provides discrete spatial snapshots without intrinsic timestamps, we define a constant user velocity of 10 m/s (36 km/h), yielding an effective sampling rate of 10 Hz. To evaluate whether the VLA pipeline can identify the approaching blockage event and generate the correct handover action \textit{before} physical degradation occurs, we instructed the agent to calculate a semantic risk score (Low, Medium, High) based on trajectory analysis relative to the bus position, as illustrated in Fig. \ref{fig:case_study_c}. The control policy was defined as:
\begin{itemize}
    \item \textit{Risk $<$ High}: Maintain current beam tracking via Tactical Controller.
    \item \textit{Risk $=$ High}: Issue proactive handover directive to Tactical Controller.
\end{itemize}

\subsubsection{Results}
Temporal analysis reveals the VLA pipeline's predictive reasoning capability, depicted in the risk transitions of Fig. \ref{fig:case_study_c}(d). At $t=0$ to $t=1.5$s, the agent detected both the vehicle and the bus but classified risk as ``Low'' due to sufficient spatial separation. As the vehicle approached the bus ($t=1.5$ to $t=1.9$s), the agent elevated risk to ``Medium''. The reasoning trace noted: ``Vehicle trajectory converging with large blocker. Gap closing at approximately 2 metres per second. Continued monitoring required.''

At $t=1.9$s the agent identifies imminent occlusion. Its trace reads \textit{``Bus dimensions block all beam paths from serving base station. Blockage inevitable within 200 milliseconds.''} The risk score reaches High, and the Planner emits a proactive handover directive for the Tactical Controller.

Ground-truth channel measurements confirm line-of-sight loss at $t=2.0$s with RSRP degradation of 12 dB. The VLA pipeline correctly identifies the blockage event and generates the appropriate handover action 100 milliseconds before physical degradation occurs, whereas a reactive baseline system (monitoring RSRP thresholds) would detect the problem only at $t=2.0$s. This demonstrates that the eBS possesses the temporal reasoning capability required for proactive beam management as it can track dynamic objects, estimate closing trajectories, and infer causal consequences for link quality, achieving physical scene understanding through zero-shot reasoning rather than task-specific supervision. Within the eBS, this predictive intelligence would be delivered as a temporally-tagged intent vector from the Semantic Planner.

\section{Discussion and Future Challenges}
\label{sec:discussion}

\subsection{Inference Latency and Model Evolution}

The eBS implements a VLA pipeline by prompting a general-purpose VLM to perceive its environment, reason about propagation physics, and generate structured control actions. This prompting-based approach offers a critical deployment advantage as it requires no paired vision-action training data, instead leveraging pre-trained world knowledge for zero-shot action generation. However, current frontier VLMs (e.g., GPT-4o, Gemini Pro) require 1--2 seconds for complex visual reasoning via cloud APIs, and action quality depends on prompt design rather than learned wireless-specific representations. Whilst the eBS architecture is designed to mask this latency for strategic planning, reducing absolute inference time would enable finer-grained temporal predictions and broader applicability to latency-critical scenarios such as the predictive handover demonstrated in Case Study C.

Two complementary directions merit investigation. First, distilling frontier VLM knowledge into compact edge-deployable models could achieve 70--80\% reasoning accuracy at 100--200 millisecond latencies, bringing inference within the Semantic Planner's 0.5--2 Hz update budget on local hardware. Second, training a dedicated wireless VLA model end-to-end, fine-tuning on datasets pairing base station camera imagery with optimal beam configurations, would replace prompt-dependent reasoning with learned wireless-specific representations, potentially improving both accuracy and speed.

Beyond VLMs, emerging world models that learn internal dynamics representations offer a natural fit for the eBS paradigm. Whilst VLMs provide semantic understanding of the current scene, world models excel at physics-aware simulation and future state prediction \cite{ding2025understanding}, and could enhance the Semantic Planner's temporal reasoning by simulating propagation consequences over longer horizons than single-frame inference allows. The two-tier eBS architecture is agnostic to the specific foundation model serving as its cognitive engine, and can readily incorporate world models as they mature.

\subsection{Edge Hardware Availability}

The case studies in this paper invoke frontier VLMs through cloud APIs, introducing network dependency and variable latency. However, recent advances in edge AI hardware are rapidly closing the gap between cloud and local inference. The NVIDIA Jetson AGX Orin already supports VLMs in the 7--20 billion parameter range at power budgets compatible with base station enclosures. Its successor, Jetson AGX Thor, delivers 7.5$\times$ greater AI compute than Orin within a 40--130 W envelope, and can run VLMs and VLA models with a time-to-first-token on the order of a few hundred milliseconds, comfortably within the Semantic Planner's 0.5--2 Hz update budget without cloud connectivity. For development and prototyping, the NVIDIA DGX Spark brings 128 GB of unified memory and 1 petaFLOP of AI performance to a desktop form factor, enabling researchers to iterate on eBS prompt design and model selection locally before edge deployment. This hardware trajectory suggests that fully self-contained eBS operation, where the entire VLA pipeline executes on co-located edge accelerators, is an engineering milestone within near-term reach rather than a long goal.

\subsection{Real-World Robustness and the Sim-to-Real Gap}

The case studies in Section \ref{sec:case_studies} validate the eBS using ray-traced digital twins and controlled vision-wireless datasets. However, operational BS imagery introduces additional challenges including lens flare, low-light conditions, weather-induced occlusions, and visual clutter. Datasets such as DeepSense 6G \cite{alkhateeb2023deepsense}, which provide synchronised camera and mmWave measurements from real urban infrastructure, offer a natural testbed for evaluating this transfer. Encouragingly, the semantic abstractions employed by the eBS (e.g., material categories, lane-level spatial regions, trajectory-based risk scores) operate at a granularity inherently more robust to pixel-level noise than supervised feature correlations, suggesting that the ``reality gap'' may prove less severe for VLM-based reasoning. Systematic validation under extreme weather and nighttime conditions remains an important direction for future work.

\subsection{Energy Efficiency and Sustainability}

A single frontier VLM inference consumes approximately 10--15 watt-hours, compared to 0.01 watt-hours for a conventional beam prediction convolutional neural network, raising scalability concerns across thousands of base stations. However, the Semantic Planner's low update rate (0.5--2 Hz versus millisecond-rate conventional processing) substantially limits total inference volume, distilled edge models offer 10--100$\times$ efficiency gains, and the 92\% beam search reduction demonstrated in Case Study B directly reduces RF pilot transmission energy at the Tactical Controller. A comprehensive lifecycle analysis should quantify whether the Planner's computational overhead is offset by these downstream savings.

\subsection{Privacy, Explainability, and Sensor Fusion}

Deploying cameras on telecommunications infrastructure raises privacy concerns. The eBS mitigates this through semantic abstraction, ensuring that only high-level scene descriptors (e.g., object categories, spatial layouts, material types) propagate beyond the edge, whilst raw imagery remains local. Future work should investigate privacy-preserving modalities (e.g., LiDAR, radar, thermal) that provide geometric information without identifying individuals. 

A complementary advantage of the VLM-based approach is interpretability. Unlike black-box neural networks, the Semantic Planner produces natural language explanations of its beam decisions, enabling operators to validate reasoning, diagnose anomalies, and progressively build trust in autonomous operation. 

Finally, the current implementation reasons over camera imagery alone, yet modern base station infrastructure increasingly carries LiDAR, radar and GPS. These modalities complement the camera rather than duplicate it. LiDAR and radar supply geometry that survives darkness and rain, and GPS anchors absolute position, whilst only the camera carries the material and semantic cues on which the eBS depends. Fusing all four through structured prompting, so that the Semantic Planner reasons over a heterogeneous scene description rather than an image alone, is the natural next step for the eBS.

\section{Conclusion}
\label{sec:conclusion}

This work introduces the eBS, a paradigm that implements a VLA pipeline for autonomous network control, where a frontier VLM perceives the physical environment, reasons about propagation physics, and generates structured control actions. A two-tier asynchronous architecture decouples strategic semantic reasoning from tactical execution, reconciling foundation model capabilities with real-time wireless requirements. By equipping base stations with the full VLA loop, comprising situated perception, causal reasoning, and autonomous action generation with natural language explainability, the eBS paradigm represents a shift from reactive signal processors and rule-following network automation to embodied intelligence, enabling wireless infrastructure that observes, understands, and autonomously optimises its interaction with the physical world.

\bibliographystyle{IEEEtran}

\bibliography{bare_jrnl}

\end{document}